\documentstyle[12pt]{article}
\newcommand{\be}{\begin{equation}}
\newcommand{\e}{\end{equation}}
\newcommand{\la}{\lambda}
\newcommand{\vi}{\varphi}
\newcommand{\is}{\int\!\!\!\!\!\!\!\mbox{$\sum$}}
\newcommand{\mf}[2]{\mbox{$\frac{#1}{#2}$}}

\begin{document}

\thispagestyle{empty}
{\parindent0em DESY 93-086 \\
June 1993}
\begin{center}
\vspace{3cm}
{\Large\bf Finite Temperature Effective Potential\\
           for the Abelian Higgs Model\\
           to the Order $e^4,\la^2$ \\}
\vspace{2cm}
Arthur Hebecker\\ \vspace{.3cm}
{\small\it Deutsches Elektronen-Synchrotron DESY, 2000 Hamburg 52, Germany\\}
\vspace{5cm}
{\bf Abstract}
\end{center}
\begin{quote}
A complete calculation of the finite temperature effective potential for the
abelian Higgs model to the order $e^4,\la^2$ is presented and the result is
expressed in terms of
physical parameters defined at zero temperature. The absence of a
linear term is verified explicitly to the given order and proven to survive to
all orders. The first order phase transition has weakened in comparison with
lower order calculation, which shows up in a considerable decrease of the
surface tension.
\end{quote}
\newpage

\renewcommand{\baselinestretch}{0.95}\small\normalsize
\section{Introduction}\vspace{-.2cm}
The electroweak phase transition \cite{KL,W,DJ} has recently attracted much
interest
due to the hope to explain the observed baryon asymmetry of the universe within
the minimal standard model \cite{KRS,FS}. Much work has also been devoted to
the phase transition in the abelian Higgs model, the simplest gauge theory with
spontaneous symmetry breaking \cite{KL2,L1}, because it is  believed to
exhibit the main features of the electroweak phase transition. An important
prerequisite to the understanding of the mechanism of the phase transition is
the knowledge of the effective potential at finite temperature \cite{KL2,Ca}.
It has already been calculated for the abelian Higgs model in \cite{BHW,Ar} to
the
order $e^3,\la^{3/2}$ and in \cite{AE} to the order $e^4,\la$. In the last
publication  the assumption $\la\sim e^3$ is made and the scalar masses $m_\vi$
and $m_\chi$ are counted as order $\la^{1/2}$, which can only be justified
close
to the critical temperature. Other approaches to the effective potential are
found in \cite{A,J}.

In this paper, assuming formally $\la\sim e^2$, a complete calculation of the
effective potential in the abelian Higgs model at finite temperature to the
order $e^4,\la^2$ is presented. Coupling constants appearing in the denominator
through infrared divergences \cite{H,BFHW} are taken into consideration and the
full dependence on the Higgs field
$\hat{\vi}$, the zero temperature vacuum expectation value $v$ and $T$ is kept.
Therefore this calculation supplies the potential given in \cite{AE} with
$\la$-corrections, which change the potential at $T_c$ and the
surface tension significantly, in spite of their numerical smallness.

The approach is based on the Dyson-Schwinger equation for the
derivative $\partial V\!/\!\partial\vi$, i.e. the tadpole diagrams are summed
\cite{W,KL1}. In section 1 the principal method of the calculation, which makes
also use of the gap equations for effective masses \cite{BHW}, is explained.

The explicit formula for $V(\vi,T)$ is given in section 2. It reproduces the
results of \cite{BHW,Ar} and \cite{AE} when taking the appropriate limits.

Section 3 addresses the problem of the linear term \cite{BHW,A,DLHLL,EQZ,BBH1}.
The expression for $V$ to the order $e^4,\la^2$ shows that a nontrivial
cancellation leads to \linebreak$\lim_{\vi\to 0}\partial V/\partial\vi=0$, i.e.
the absence of a linear term. It is proven that this feature survives to all
orders of perturbation theory.

To get rid of the arbitrary scale $\bar{\mu}$, which is introduced when
regularizing the theory in the dimensional scheme, a transition to physical
parameters is performed in section 4. However this finite renormalization at
zero temperature proves to be not very important numerically.

The numerical analysis in the last section concentrates on the surface tension
\cite{C,L}, because this easily accessible parameter
gives a first characteristic of the strength of the first order phase
transition, which is important for the generation of the baryon asymmetry in
the standard model. The surface tension is found to be generally smaller than
the lower order calculation \cite{BHW} suggests. This decrease is dramatic for
large $\la$, but in contrast to the $e^4,\la$-potential \cite{AE}, the
potential discussed here does not show a complete change to a second order
phase transition in the considered domain of $\la$.

\renewcommand{\baselinestretch}{1.00}\small\normalsize

\section{Calculation of the Effective Potential using \hspace{3cm}\mbox{}
Dyson-Schwinger Equations}

\subsection{Explanation of the Method in $\la\vi^4$-Theory}
After describing the calculation in some detail in simple $\la\vi^4$-theory
the extension to the abelian Higgs model is shown to be straightforward.

The euclidean Lagrangian has the form
    \be {\cal L}=-\frac{1}{2}(\partial\vi)^2+\frac{\nu}{2}\vi^2-
    \frac{\la}{4}\vi^4, \e
where $\nu=\la v^2$ is counted as order $\la$.
Using the familiar zero temperature technique of Dyson-Schwinger equations
\cite{Riv} it is easy to obtain the relation \\[0.8cm]
    \be -\frac{\partial}{\partial\hat{\vi}}(V-V_{tree})=A+B=\hspace{3.8cm}+
    \hspace{4.2cm}.\label{0}\e\\[0.8cm]
Here the 3-vertex in the first term arises from the shift $\vi\rightarrow\vi+
\hat{\vi}$. The two different sorts of blobs symbolize the full propagator and
the full 3-vertex.

The next step is to investigate the first term
    \be A=-\la\hat{\vi}\is\frac{dk}{k^2+m_{tree}^2+\Pi(k)}\quad,\label{1}\e
where the tree level mass square $m_{tree}^2=\la\hat{\vi}^2-\nu$ is assumed to
be of order $\la$. For a calculation to order $\la^2$ it is sufficient
to know $\Pi(k_0\neq 0,\vec{k})$ and $\Pi(0,\vec{y}m)$ to order $\la$ and
$\la^{3/2}$ respectively. Here the need for different treatment
of the contribution with zero Matsubara frequency can be understood by
performing the substitution $\vec{k}\to\vec{y}m$ in the integral.
( See \cite{BHW,H,BFHW} for the correct way of counting
the order of infrared contributions here and below. ) Therefore in the
Dyson-Schwinger equations for $\Pi(k)$ the vertex correction can be neglected
\cite{BHW} :\\[1.2cm]
    \be -\Pi(k)=-(\Pi_a(k)+\Pi_b(k))=\hspace{3.2cm}+\hspace{5.0cm}.\e\\[0.7cm]
With the definitions
    \begin{eqnarray}
    m_3^2=m_{tree}^2+\Pi_a(0)|_{\la^{3/2}}+\Pi_b(0)|_{\la}\quad & , & \quad
    \Pi_{02}=\Pi_b(0)|_\la\quad,\label{3}\\
    \Pi_{03}=\Pi_b(0)|_{\la^{3/2}}-\Pi_b(0)|_\la\qquad\qquad\quad & , & \quad
    \Pi_1(k)=\Pi(k)-\Pi(0)\quad,\nonumber\end{eqnarray}
where the powers of $\la$ symbolize the accuracy to which a certain term has to
be calculated, the following relation holds:
    \be m_{tree}^2+\Pi(k)=m_3^2+\Pi_{03}+\Pi_1(k)+\mbox{O}(\la^2) .\label{4}\e
Of course $\Pi_{02}$ vanishes in $\la\vi^4$-theory, but keeping this term
formally
makes the extension to the abelian gauge theory with power counting rule $\la
\sim e^2$ more simple.
After substituting (\ref{4}) into equation (\ref{1}) the integrand is expanded
neglecting contributions of order higher than $\la^2$, which results in the
formula
    \be A=-\la\hat{\vi}\is dk\left(\frac{1}{k^2+m_3^2}-\frac{\Pi_{03}+\Pi_1(k)}
    {(k^2+m^2)^2}\right)=\label{2}\e
    \[ =-\la\hat{\vi}\is dk\left(\frac{1}{k^2+m_3^2}+\frac{\Pi_{02}}
    {(k^2+m^2)^2}-\frac{\Pi_{02}+\Pi_{03}+\Pi_1(k)}{(k^2+m^2)^2}\right). \]
Here $m^2=m_3^2|_\la$ is the leading order mass
term including the temperature correction. Observing that in term $B$ of
equation (\ref{0}) the vertex need not be corrected to obtain the full
$\la^2$-result it becomes obvious that the third term in the right hand side of
(\ref{2}) together with $B$ is equal to the derivative of the
two-loop diagram\\[.6cm]
    \be \frac{\partial}{\partial\vi}\{V_s\}=\frac{\partial}{\partial\vi}\bigg\{
    \hspace{3cm}\bigg\},\label{5}\e\\[.6cm]
with leading order mass corrections in the propagators.
So the final expression for the potential is
    \be V=V_{tree}+\int^\vi d\vi'\la\vi'\is dk\left(\frac{1}{k^2+m_3^2}+
    \frac{\Pi_{02}}{(k^2+m^2)^2}\right)+V_s.\e
A similar way of combining the different contributions to $V$ has been
considered in \cite{BBH}.

\subsection{Extension to the Abelian Higgs Model}
The abelian Higgs model in Landau gauge includes, from the topological point of
view, exactly the same graphs, because it does also contain 3-and 4-vertices.
Consider the euclidian Lagrangian
    \be {\cal L}=-\frac{1}{4}F_{\mu\nu}F_{\mu\nu}-|D_\mu\Phi|^2+\nu|\Phi|^2-\la
    |\Phi|^4,\quad\Phi=\frac{1}{\sqrt{2}}(\hat{\vi}+\vi+i\chi).\e
\renewcommand{\baselinestretch}{0.97}\small\normalsize
Applying the formal power counting rule $\la\sim e^2$ the main difference from
the purely scalar model lies
in the low order of the vector-scalar-scalar vertex.
This manifests itself through the appearance of nonzero terms $\Pi_{02,T},
\Pi_{02,L}$ with $T,L$ referring to contributions to the transverse and
longitudinal parts of the vector propagator \cite{BHW,K} from the diagram in
fig. 1.\\[0.6cm]
\hspace*{\fill}Fig. 1\\[0.6cm]
But this feature has already been accounted for in the last subsection by
keeping $\Pi_{02}$. Therefore the contributions to the potential can be
classified analogously to the purely scalar case described above. This can also
be shown starting directly from the Dyson-Schwinger equations of the abelian
Higgs model. Of course at each step it is necessary to convince oneself that no
contributions of order $\la^2,e^4$ are lost. The final result has exactly the
same structure as the potential of the last subsection, but it involves more
terms due to the particle content of the theory.
    \be V=V_{tree}+V_1+\cdots+V_8+V_{3,\Pi}+V_{4,\Pi} \e\\[-.1cm]
Here\\[-.1cm]
    \be V_1=\int^\vi d\vi'\la\vi'I(m_{\chi,3})\quad,\quad
       V_2=\int^\vi d\vi'3\la\vi'I(m_{\vi,3})\quad,\e
    \[ V_3=\int^\vi d\vi'(2-2\epsilon)e^2\vi'I(m_{T,3})\quad,\quad
       V_4=\int^\vi d\vi'e^2\vi'I(m_{L,3})\]
with the standard temperature integral \cite{DJ}\\[-.1cm]
    \be I(m)=\is dk\frac{1}{k^2+m^2} \e\\[-.1cm]
to be evaluated in $n-1=3-2\epsilon$ dimensions. The masses are defined in
analogy with equation (\ref{3}) by the appropriate zero momentum parts of the
Dyson-Schwinger equations, which have to be iterated once to obtain the result
to order $e^3,\la^{3/2}$ \cite{BHW}. Corrections from the momentum dependent
part of the Dyson-Schwinger equations are needed for the gauge boson only:
    \be V_{3,\Pi}\!=\!\!\int^\vi \!d\vi'(2\!-\!2\epsilon)e^2\vi'\is dk\frac{
    \Pi_{02,T}}{(k^2+m_T^2)^2}\;,\;V_{4,\Pi}\!=\!\!\int^\vi\! d\vi'e^2\vi'\is
    dk\frac{\Pi_{02,T}}{(k^2+m_L^2)^2}.\e
\renewcommand{\baselinestretch}{1.00}\small\normalsize
Finally $V_5$ through $V_8$ represent the diagrams of figure 2.\\[1.5cm]
\hspace*{\fill} Fig. 2\\[1.5cm]\mbox{}
{\samepage\parindent0em Furthermore the contributions generated by the
counterterm Lagrangian have to be added.}

\section{Explicit Result to Order $e^4,\la^2$}
The calculations needed for the temperature dependent masses to order
$e^3,\la^{3/2}$ can essentially
be taken from \cite{BHW}. Differences arise because the loop with two
propagators does only contribute in lowest order due to definition (\ref{3}).
Also it is necessary to keep the $\epsilon$-dependence in leading order.
The integrals in $V_5$ through $V_8$ have already been done in \cite{AE,P}
and wont be given here explicitly. After adding all the terms up, which proves
to be rather laborious, the final
result, using the $\overline{\mbox{MS}}$-scheme, reads
\begin{eqnarray}
    V(\vi)&\!\!\!\!=&\!\!\!\!\frac{\vi^2}{\beta^2}\Bigg[-\frac{\beta^2\nu}{2}
    +\frac{\la}{6}+\frac{e^2}{8}+\frac{e^4}{64\pi^2}\left(-\mf{11}{3}
    \ln{\bar{\mu}^2\beta^2}+\mf{2}{3}c+3c_2-\mf{13}{9}\right)\label{6}\\
    &&\nonumber\\
    &&\!\!\!\!\!\!\!\!\!\!\!\!
    +\frac{e^4}{16\pi^2}\left(\ln{\mf{2m_L+m_\vi}{2m+m_\vi}}+
    \mf{1}{2}\ln{
    \mf{m+m_\vi+m_\chi}{m+m_\vi}}+3\ln{\mf{\beta(2m+m_\vi)}{3}}\right)
    +\frac{\beta^2\nu\la}{8\pi^2}\left(\ln{\bar{\mu}^2\beta^2}-c+\mf{3}
    {2}\right)\nonumber\\&&\nonumber\\
    &&\!\!\!\!\!\!\!\!\!\!\!\!
    +\frac{\la e^2}{16\pi^2}\left(\mf{1}{2}\ln{\bar{\mu}^2\beta^2}+\mf{1}{2}c-
    c_2+\mf{1}{3}-4\ln{\mf{\beta(m+m_\vi+m_\chi)}{3}}-3\ln{\mf{(2m+m_\vi)}{(m+
    m_\vi)}}\right)\nonumber\\&&\nonumber\\
    &&\!\!\!\!\!\!\!\!\!\!\!\!
    +\frac{\la^2}{16\pi^2}\left(-\mf{5}{3}\ln{\bar{\mu}^2\beta^2}\!+\!\mf{2}
    {3}c\!+\!c_2\!+\!2\!+\!\ln{\mf{\beta(m_\vi+2m_\chi)}{3}}\!+\!3\ln{\beta
    m_\vi}\!+\!2\ln{\mf{m+m_\vi+m_\chi}{m_\vi+m_\chi}}\right)\Bigg]\nonumber\\
    &&\nonumber\\
    &&\!\!\!\!\!\!\!\!\!\!\!\!
    +\vi^4\Big[\;\;\frac{\la}{4}
    +\frac{1}{64\pi^2}(10\la^2+3e^4)(c-\mf{3}{2}
    -\ln{\bar{\mu}^2\beta^2})+\frac{e^4}{32\pi^2}\Big]\nonumber\\&&\nonumber\\
    &&\!\!\!\!\!\!\!\!\!\!\!\!
    +\frac{m_\vi^4}{64\pi^2\beta^2\vi^2}\ln{\mf{m_\vi(
    2m+m_\vi)}{(m+m_\vi)^2}}-\frac{M_0^2e^2}{16\pi^2\beta^2}\left(2\ln{\mf{
    \beta(m+m_\vi+m_\chi)}{3}}+\ln{\mf{2m+m_\vi}{m+m_\vi}}\right)\nonumber\\
    &&\nonumber\\
    &&\!\!\!\!\!\!\!\!\!\!\!\!
    +\frac{1}{32\pi^2\beta^2}\!\left[(e^2\!+\!\la)m_\vi m_\chi\!+\!e^2
    m_L(m_\vi\!+\!m_\chi)\!+\!2\la m(m_\chi\!-\!m_\vi)\!+\!e^2m(m_\chi
    \!+\!2m_\vi)\right]\nonumber\\&&\nonumber\\
    &&\!\!\!\!\!\!\!\!\!\!\!\!
    -\frac{1}{12\pi\beta}(m_\vi^3+m_\chi^3+2m^3+m_L^3).\nonumber\end{eqnarray}
Here the lowest order masses are
    \be m_\vi^2=3\la\vi^2+M_0^2\quad,\quad m_\chi^2=\la\vi^2+M_0^2\quad,\quad
    M_0^2=-\nu+\frac{4\la+3e^2}{12\beta^2}\quad,\e
    \[  m^2=e^2\vi^2\quad,\quad m_L^2=e^2\vi^2+\frac{e^2}{3\beta^2}\]
and the constants $c$ and $c_2$ arise from the temperature integral $I(m)$
(see \cite{DJ}) and the scalar two loop integral in (\ref{5}) calculated in
\cite{P}
    \be c=\mf{3}{2}+2\ln{4\pi}-2\gamma\approx 5.4076\quad,\quad c_2\approx
    3.3025\quad.\e
Dropping terms of order $e^4$ (with power counting rule $\la\sim e^2$ ) the
effective potential from \cite{BHW} is recovered. When changing the power
counting to $\la\sim e^3$ and $m_\vi^2,m_\chi^2\sim\la$ the result of Arnold
and Espinosa
\cite{AE} is obtained after dropping terms of order higher than $e^4$.
It should be noted that the term in (\ref{6}) containing an explicit
$1/\vi^2$-factor does not show singular behaviour near $\vi=0$ because of the
logarithm, which decreases fast enough.

\section{Absence of a linear Term}
At first sight the contribution to the potential (\ref{6}) proportional to
$m(m_\chi+2m_\vi)$ seems to
produce a linear behaviour for small $\vi$. But in this limit
the logarithmic terms have to be expanded at the point $\vi=0$, which results
in
linear terms exactly cancelling the one mentioned above. In fact this feature,
already discussed by several authors \cite{BHW,A,DLHLL,EQZ} starting from lower
order calculations, can be shown to
survive to all orders of small couplings perturbation theory :
    \[ \lim_{\vi\to 0}\frac{\partial V}{\partial\vi}=0 \mbox{ to all orders in
    $e$ and $\la$.}\]
This can be proven in the following way (compare the argument in
\cite{DLHLL}) : Global U(1)-symmetry implies
    \be \frac{1}{\vi}\frac{\partial V}{\partial\vi}=m_\chi^2(q^2=0)\quad. \e
Obviously it suffices to show, that the self energy $\Pi_\chi(q^2=0)$ is finite
for $\vi\to 0$. Above the barrier temperature singularities can only arise from
the transverse gauge boson propagator due to the temperature masses of $\chi$
and $\vi$. Therefore diagrams of the kind shown in fig. 3 have to be
investigated. Here the wavy lines symbolize full vector propagators and the
blobs are vertices without internal vector lines, meaning the sum of all
diagrams built from scalar propagators with the correct number of external
vector lines.\pagebreak

\vspace*{1.5cm}

\hspace*{\fill} Fig. 3\\[1.5cm]
If $\vi=0$, a gauge covariance argument, completely analogous to the zero
temperature case, shows for the vertices with external vector lines only, that
    \be \Gamma^{2n}_{\alpha\beta\ldots\mu\nu}\sim|\vec{k_1}|\ldots
    |\vec{k_{2n}}|\quad\mbox{for small}\;|\vec{k_i}|\quad\mbox{and}\quad
    k^0_i=0,\;\; \alpha\beta\ldots\mu\nu\in\{1,2,3\}. \e
If $\vi\neq 0$, diagrams not covered by the previous gauge covariance argument
because of explicit $\vi$-factors at the vertices have to be added to $\Gamma
^{2n}$. They however vanish not slower than $\vi^2$ in the limit
$\vi\to 0$, which is clear from the fact that the unbroken theory has no
vertices with an odd number of scalar lines.
Therefore in the case $|\vec{k}|\sim\vi$ ($\vi$ being the natural
infrared cutoff introduced by the transverse vector mass $e\vi$) the sum of
both
contributions, i.e. the complete vertex $\Gamma^{2n}$, can be counted as
$\vi^2$ when searching for small-$\vi$ singularities.

Consider the most dangerous lowest power of $\vi$ stemming from the maximal
infrared divergence, which is obtained by setting $k^0=0$ for all transverse
vector propagators. It can be calculated by scaling the loop momenta according
to $\vec{k}\to\vec{y}\vi$. Counting the
vector vertices as $\vi^2$ the following formula for the minimal overall power
of $\vi$ is obtained (compare the argumentation in appendix A of \cite{BFHW}) :
\be n_\vi=3L-2I+2(V-2)+2\quad. \label{77}\e
Here $L,I$ and $V$ denote the number of vector loops, vector propagators and
full vertices, symbolized by blobs in fig. 3, respectively. The last term +2
follows from a closer look at the contribution of the full vertices with
external $\chi$-lines: Fig. 3 shows examples of the two different structures
to be investigated.
If there are two such "$\chi$-vertices", each will contribute a factor $\vi$,
resulting in the correction +2. If there is only one with two external
$\chi$-lines, it may have no explicit $\vi$-factor. In the latter case however
the contribution of the vector vertices $2(V-2)$ has to be replaced by
$2(V-1)$,
which again corresponds to a correction +2. Therefore equation (\ref{77}) is
valid in the general case. Now using the well-known formula $V+L-I=1$ it
follows
immediately that
\be n_\vi=L\ge 0, \e
or equivalently: There is no divergence for $\vi\to 0$. If not all
of the vector propagators are infrared divergent, the vertices connected by
"heavy" lines may be formally fused. Now repetition of the above argument
leads again to the desired result thus completing the proof.

\section{Transition to physical Parameters}
To get rid of the arbitrary scale $\bar{\mu}$ the potential is rewritten in
terms of physical parameters defined at zero temperature. Such parameters are
the Higgs and vector masses and the vacuum
expectation value of the Higgs field. To stay closer to the previous notation
they can be expressed through new coupling constants
 $\bar{\la}$ and $\bar{e}$, defined by
    \be m^2_{\vi,phys}=2\bar{\la}v^2,\quad m^2_{phys}=\bar{e}^2v^2,\quad
    \frac{\partial V}{\partial\vi_{phys}}\Bigg|_{\vi_{phys}=v}=0.\label{7}\e
Effectively a finite renormalization of the form
    \be \vi^2=\vi_{phys}^2(1+c),\quad \la=\bar{\la}+\delta\bar{\la},\e
    \[ e^2=\bar{e}^2+\delta \bar{e}^2,\quad \nu=\bar{\la}v^2+\delta
    \nu_{phys}\]
has to be performed. The physical $\vi$-propagator is
    \be \frac{1}{(1+c)(q^2-m_\vi^2-\Pi_\vi(q^2))}, \e
therefore the on-shell definitions of the new parameters follow from
    \be c=\frac{\partial}{\partial q^2}\mbox{Re}\;\Pi_\vi(q^2)\Big|_{q^2=m_
    {\vi,phys}^2},\e
    \[ m_\vi^2+\mbox{Re}\;\Pi_\vi(m_{\vi,phys}^2)=m_{\vi,phys}^2,\quad
    m^2+\mbox{Re}\;\Pi(m_{phys}^2)=m_{phys}^2 \]
together with the last equality in (\ref{7}). The zero temperature effective
potential to the order $\la^2,e^4$ needed here is given by
    \begin{eqnarray} V & = & -\frac{\nu}{2}\vi^2+\frac{\la}{4}\vi^4-\frac{m_
    \chi^4}{64\pi^2}\left(\mf{3}{2}+\ln{\mf{\bar{\mu}^2}{m_\chi^2}}\right)\\
    &&-\frac{m_\vi^4}{64\pi^2}\left(\mf{3}{2}+\ln{\mf{\bar{\mu}^2}{m_\vi^2}}
    \right)-\frac{3m^4}{64\pi^2}\left(\mf{5}{6}+\ln{\mf{\bar{\mu}^2}{m^2}}
    \right).\nonumber\end{eqnarray}
Contributions to the self energy corrections $\Pi_\vi(q^2)$ and $\Pi(q^2)$ come
from the usual zero temperature one loop diagrams. In view of the principal
features of the potential considered in this paper the numerical effect of the
performed finite renormalization is not very important (see fig. 5 in the last
section). Therefore the complete formula for $V$, which is easy to obtain, is
not given here explicitly.

\section{Numerical Results and Discussion}
In the previous sections a complete calculation of the finite temperature
effective potential to the order $e^4,\la^2$ has been performed, including the
transition to physical parameters defined at zero temperature. The gauge
coupling is chosen to be
$e=0.3$ and the influence of $\la$, which corresponds to the Higgs mass, is
investigated. At first sight the calculated potential seems to ensure a first
order phase transition in a wide range of the parameter $\la$, but
reliability of perturbation theory has to be questioned. This becomes obvious
from fig. 4, where different approximations of the effective potential at their
respective critical temperatures are shown. (Here and below the dimensionful
quantities are given in units of $v$ and its powers.)
Some insight can be
gained from a comparison with the results to order $e^3,\la^{3/2}$ obtained
in \cite{BHW,Ar}. A reasonable physical quantity to be calculated from both
potentials is the surface tension \cite{C,L},
which can be seen as a measure of the strength of the phase transition :
\be \sigma=\int_0^{\vi_+}d\vi\sqrt{2V(\vi,T_c)}, \e
where $\vi_+$ is the position of the second degenerate minimum of $V$ and the
potential is
normalized to ensure $V(\vi=0)=0$. The numerical results are shown in fig. 5.
It includes besides the surface tension from potentials to the order
$e^3,\la^{3/2}$ \cite{BHW} and to the order $e^4,\la^2$ also the results
calculated from
a potential to order $e^4,\la$ \cite{AE}, where according to the power counting
rule
$\la\sim m_\vi^2\sim m_\chi^2\sim e^3$ all terms of order higher than $e^4$
have
been neglected. Obviously the shift introduced by the transition to zero
temperature physical parameters is not important for the present discussion.

The fact that perturbation theory is not reliable for large $\la$, already
stressed in \cite{BHW}, can be clearly read off from fig. 5 ($\sigma$ changes
by
an order of magnitude when adding the last term in the perturbation series).
Somewhat surprisingly the perturbation series cannot be trusted too much for
small $\la$ either. Here $\sigma$ decreases by a factor of $\sim 2$ at least.
Two different higher order terms are mainly responsible this for this change.
In the region of extremely small $\la$ it is essentially the $e^4\vi^4$ term
(see equation (\ref{6})), which cannot be viewed as a small correction to the
tree level term $\la\vi^4/4$. This enhancement of the $\vi^4$-term is a
temperature effect and therefore cannot be removed by zero temperature
renormalization. For $m_\vi^2/m^2\sim 0.4$ the logarithmic mass dependence of
the
$e^4\vi^2$ term seems to be more important. The great influence of this
numerically small contribution can be understood by recalling that at the
critical temperature the leading order $\vi^2$-terms essentially cancel and
that a $\vi$-dependence in a coefficient of $\vi^2/\beta^2$ cannot be absorbed
in a correction \linebreak of $T_c$.

It is interesting to compare the above discussion with another method to
investigate the reliability of the perturbation series: Applying the
$\xi$-conditions introduced in \cite{BHW} in the more restrictive form
of \cite{BFHW} (i.e. using the higher order expression for the masses)
the region of reliability $m_\vi^2/m^2\stackrel{<}{\sim}0.1$ is obtained for
$\xi$=2. For the largest permissible Higgs mass fig. 5 suggests $\Delta\sigma/
\sigma=0.57$ which signals the breakdown of the perturbation series.
This error does not decrease for smaller Higgs masses, in contrast to the
standard model calculation in lower order of \cite{BFHW}, due to the
shift of $\la$ through an $e^4$-term discussed above. However, this enhancement
of the $\vi^4$-contribution, which is invisible in the $\xi$-conditions, does
not threaten the first order of the phase transition. Therefore perturbation
theory seems to ensure a first order phase transition for small $\la$ at least,
in spite of the still unknown exact value of the surface tension.

The potential to the order $e^4,\la$ from \cite{AE} does not give rise to
a first order phase transition for $\la\stackrel{>}{\sim}0.01$. The value of
$\sigma$ calculated using this potential differs significantly from the
result presented here for $\la\stackrel{>}{\sim}0.007$. This is partially
due to contributions of the form $e^4\vi^2\ln(m+m_\vi)$ and the like,
already mentioned above. Counting $m_\vi$ as higher order correction results
in the $\ln m$-contributions found in the potential from \cite{AE}, which is
obviously a significant change for small $\vi$.

Fig. 6 shows the vacuum expectation value in the asymmetric phase at $T_c$.
This parameter does not reflect the dramatic change of the surface tension
by higher order corrections.

It would be interesting to extend this approach
to the standard model, which seems to be straightforward, and to try to
estimate the influence of expected corrections beyond $e^4,\la^2$.

I am most grateful to W. Buchm\"uller, who suggested this investigation, for
\linebreak
continuous support and encouragement. Also many helpful discussions with
\linebreak
D. B\"odeker, Z. Fodor, T. Helbig and H. Kohrs have to be acknowledged.
\\[2cm]
{\bf Fig.4} Different approximations of the effective potential plotted at
      their respective critical temperatures with $\la$=0.01 ( the
      $e^4,\la$-potential is a result of \cite{AE} )\\[.3cm]
{\bf Fig.5} Dependence of the surface tensions calculated from the different
      potentials on the zero temperature mass relation $m_\vi^2/m^2=2\la/e^2$
      with $e=0.3$\\[.3cm]
{\bf Fig.6} Dependence of the position of the second minimum of different
      potentials on the zero temperature mass relation $m_\vi^2/m^2=2\la/e^2$
      with $e=0.3$\\[2cm]

\end{document}